\documentclass[11pt]{article}\usepackage[]{graphicx}\usepackage[]{color}
\makeatletter
\def\maxwidth{ %
  \ifdim\Gin@nat@width>\linewidth
    \linewidth
  \else
    \Gin@nat@width
  \fi
}
\makeatother

\usepackage{alltt}
\usepackage{bm}
\usepackage{geometry}
\usepackage{url}
\usepackage[american]{babel}
\usepackage{graphicx}
\usepackage{rotating}
\usepackage{tabularx}
\usepackage{amssymb,amsmath}
\usepackage{ifxetex,ifluatex}
\usepackage{fixltx2e} 

\usepackage[round,authoryear]{natbib}
\usepackage{tipa}

\IfFileExists{upquote.sty}{\usepackage{upquote}}{}
\ifnum 0\ifxetex 1\fi\ifluatex 1\fi=0 
  \usepackage[utf8]{inputenc}
\else 
  \ifxetex
    \usepackage{mathspec}
    \usepackage{xltxtra,xunicode}
  \else
    \usepackage{fontspec}
  \fi
  \defaultfontfeatures{Mapping=tex-text,Scale=MatchLowercase}
  
\fi

\IfFileExists{upquote.sty}{\usepackage{upquote}}{}

\title{
\rule[0.4cm]{\textwidth}{2pt}
{\bf Permutation-based cluster size correction for voxel-based lesion-symptom mapping}
\rule{\textwidth}{2pt} 
}

\begin{document}
\maketitle
\begin{center}
Daniel Mirman$^{a,b,}$\footnote{Address correspondence to Daniel Mirman, dan@danmirman.org} \\
Jon-Frederick Landrigan$^a$ \\
Spiro Kokolis$^a$  \\
Sean Verillo$^a$ \\
Casey Ferrara$^b$ \\ \ \\
$^a$ Drexel University, Philadelphia, PA, USA\\
$^b$ Moss Rehabilitation Research Institute, Elkins Park, PA, USA \\
\end{center}

\begin{center}
  {\bf Abstract}
\end{center}

\noindent Voxel-based lesion-symptom mapping (VLSM) is a major method for studying brain-behavior relationships that leverages modern neuroimaging analysis techniques to build on the classic approach of examining the relationship between location of brain damage and cognitive deficits. Testing an association between deficit severity and lesion status in each voxel involves very many individual tests and requires statistical correction for multiple comparisons. Several strategies have been adapted from analysis of functional neuroimaging data, though VLSM faces a more difficult trade-off between avoiding false positives and statistical power (missing true effects). One such strategy is using permutation to non-parametrically determine a null distribution of cluster sizes, which is then used to establish a minimum cluster size threshold. This strategy is intuitively appealing because it respects the necessary spatial contiguity of stroke lesions and connects with the typical cluster-based interpretation of VLSM results. We evaluated this strategy for detecting true lesion-symptom relations using simulated deficit scores based on percent damage to defined brain regions (BA 45 and BA 39) in a sample of 124 individuals with left hemisphere stroke. Even under the most conservative settings tested here, the region identified by VLSM with cluster size correction systematically extended well beyond the true region. As a result, this strategy appears to be effective for ruling out situations with no true lesion-symptom relations, but the spatial contiguity of stroke lesions may cause identified lesion-symptom relations to extend beyond their true regions. \\


\section{Introduction}

Identifying relationships between location of brain damage and cognitive deficits is a foundational method in cognitive neuroscience, tracing its history at least to the behavioral neurologists of the mid-19th century (e.g., \citealp{Lichtheim1885}). Those early studies were based on individual case studies and, as data accumulated, researchers used lesion overlays to identify the locations where damage consistently produced deficits of interest. Recent advances in neuroimaging technology have allowed much finer-grained analyses at the level of individual voxels \citep{Bates2003, Rorden2004}. In voxel-based lesion-symptom mapping (VLSM), an association between deficit severity and lesion status (lesioned vs. not lesioned) is tested in each voxel, producing a statistical map of the strength of relationship of lesion status and deficit. However, this map is the result of very many individual tests, typically thousands or tens of thousands of voxels are included in the region of interest. 

This large number of tests requires some kind of statistical correction for multiple comparisons. Several strategies have been proposed, often by adaptation from analysis of functional neuroimaging (e.g., fMRI). These strategies generally fall into two families. The first seeks to control family-wise error rate (FWER), which is the probability of making one or more false positive (Type 1) errors among entire set of tests. One classic method of FWER correction is the Bonferroni correction, which is now rarely used for neuroimaging data because it assumes independence between tests and is overly conservative when the comparisons are not independent (which is likely for neighboring voxels). Another strategy is to build a null distribution of the test statistic based on permutations of real data and to select only voxels where the true analysis stands out from the permutation-based null distribution (e.g., \citealp{Kimberg2007}).

A key aspect of FWER correction is that it controls the probability of making one or more false positive errors. This may be unnecessarily conservative because VLSM interpretation never depends on a single voxel so a single false positive voxel cannot be responsible for an invalid inference about lesion-symptom relations. This statistical conservatism comes with a substantial price: VLSM analysis is based on a single data point per participant (each participant only has one lesion and only one deficit profile) and sample sizes are often limited by the practical barriers to recruiting large numbers of participants with the targeted neurogenic deficits and the willingness and ability to participate in behavioral protocols and neuroimaging.

Because VLSM interpretation depends on the overall pattern rather than single voxels, a second family of correction methods focuses on controlling errors at the level of the overall pattern of lesion-symptom relations while allowing one or more individual voxels to be false positives. One such technique is False Discovery Rate (FDR), which quantifies the proportion of above-threshold results that can be expected to be false positives \citep{Genovese2002}. That is, at FDR threshold $q = 0.05$, 5\% of above-threshold voxels are expected to be false positives, which is likely to be substantially more than one voxel (for a clear description see \citealp{Bennett2009}). FDR is widely used for analysis of functional neuroimaging data and has been used for VLSM. We have encountered informal criticism that FDR is inappropriate for VLSM, but we are not aware of any formal analysis supporting these criticisms.

Another strategy for controlling pattern-level error is to focus on size of supra-threshold voxel clusters rather than individual voxels. Setting a minimum cluster size is a common “clean-up” step in neuroimaging data analysis; the addition of a principled strategy for selecting the minimum cluster size is the critical component that turns this into a statistical correction method. This principled strategy can be based on permutation and proceeds as follows: (1) permute behavioral data and conduct VLSM analysis, (2) apply pre-set voxel-wise threshold (e.g., $p < 0.0001$), (3) compute sizes of supra-threshold voxel clusters (e.g., \citealp{Nichols2002}). These three steps are repeated many times to build up a null distribution of supra-threshold cluster sizes. Clusters from the original (true) VLSM analysis that are larger than 95\% of the null distribution of cluster sizes are taken to reflect true lesion-symptom associations (for examples of application to VLSM see \citealp{Pillay2014, Mirman2015a}). That is, there are two thresholding steps: in the first step, a pre-set voxel-wise p-threshold is applied; in the second step, a permutation-based cluster size threshold is applied. 

This strategy has two intuitively appealing properties. First, stroke produces spatially contiguous lesions, resulting in high spatial correlations for lesion status of neighboring voxels. Using permutation to determine a null distribution of cluster sizes, intuitively conrols for this spatial correlation and produces a minimum cluster size threshold that should not be observed by chance. Second, interpretation of VLSM (and other neuroimaging) results typically focuses on clusters, so by correcting at the cluster level rather than the voxel level, this statistical strategy is more closely related to the interpretion strategy. In the following analyses we examine this permutation-based cluster-size correction strategy for detecting true lesion-symptom relations.

\section{Methods}

\subsection{Data}

\begin{figure}[ht]
\centering
\includegraphics[width=1.0\textwidth]{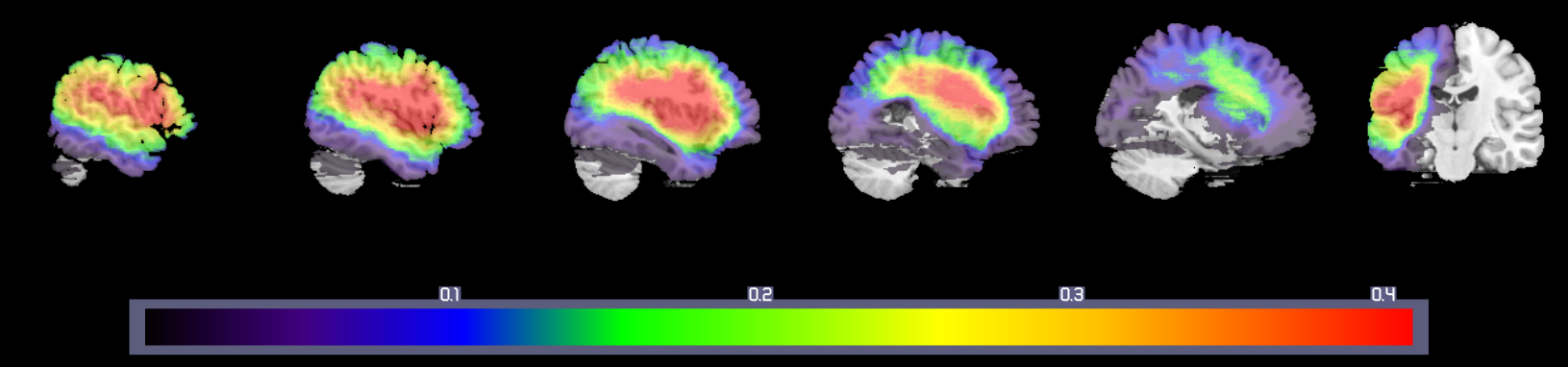}
\caption{Lesion overlap map for 124 left hemisphere stroke cases included in the present analyses. Hotter colors indicate that a larger proportion of the participant sample had lesions in that area.}
\label{fig:LesionOverlap}
\end{figure}

The lesion maps were from 124 participants with aphasia following left hemisphere stroke confirmed by computed tomography (CT) or magnetic resonance imaging (MRI). The structural data were based on 108 research scans (65 MRI and 43 CT) and 16 clinical scans (5 MRI and 11 CT). Lesions imaged with MRI were manually segmented on the structural image by a trained technician and reviewed by an experienced neurologist, then registered first to a custom template constructed from images acquired on the same scanner, and then from this intermediate template to the Montreal Neurological Institute space ‘Colin27’ volume. Lesions imaged with CT were drawn by the experienced neurologist directly onto the Colin27 volume, after rotating (pitch only) the template to approximate the slice plane of the patient's scan. Figure \ref{fig:LesionOverlap} shows the lesion overlap map for these 124 lesion maps, which have been used in VLSM analyses reported elsewhere \citep{Mirman2015a, Mirman2015b} and are part of a larger, ongoing project investigating the anatomical basis of psycholinguistic deficits in post-acute aphasia\footnote{That project has been funded by National Institutes of Health grant R01DC000191 to Myrna F. Schwartz and we are grateful to Dr. Schwartz and her team for sharing these data with us and making these analyses possible.}.

In order to have a deficit score with a known neural correlate, we calculated the percent damage in two brain regions that are widely-studied and frequently damaged in middle cerebral artery stroke aphasia: BA 45 and BA 39 (see Figure \ref{fig:LesionOverlap}). An effective statistical correction strategy should approximately identify these areas (i.e., damage in BA 45 is the ``neural correlate" of percent damage in BA 45) and the permuted data will provide additional insight into the method’s ability to reject false positives. 

\subsection{Analysis Strategy}

For each of the deficit (percent damage) scores, we conducted a basic VLSM, applied a pre-set threshold, then calculated the sizes of supra-threshold voxel clusters. We then repeated this analysis 1000 times, permuting the deficit scores for each repetition to create a random association between the scores and lesion profiles. The cluster sizes from the permutations were used to set a 95\% threshold (i.e., larger than 95\% of permutation-based clusters) for the original VLSM data. These analyses were implemented using SPM8 in Matlab and cluster sizes were computed using the {\sc bwconncomp} function from the Image Processing Toolbox. 

Six different pre-set thresholds were tested within the same set of 1000 permutations: 0.05, 0.01, 0.005, 0.001, 0.0005, 0.0001. This covers the range from the most permissive threshold (0.05) to a reasonably conservative threshold (0.0001) for initially identifying voxels for subsequent cluster size correction. The more permissive thresholds will allow more voxels into the cluster size calculation, which should produce larger clusters. Therefore, there should be a positive correlation between the pre-set p-threshold and the permutation-based cluster size threshold. This positive correlation is an inverse strictness relationship: more permissive p-thresholds produce more conservative cluster size thresholds. One motivation for this study was to examine how one might balance these inversely related factors for optimal VLSM interpretation and inference.

\section{Results}

\subsection{Analysis 1: All Clusters}

\begin{figure}[ht]
\centering
\includegraphics[]{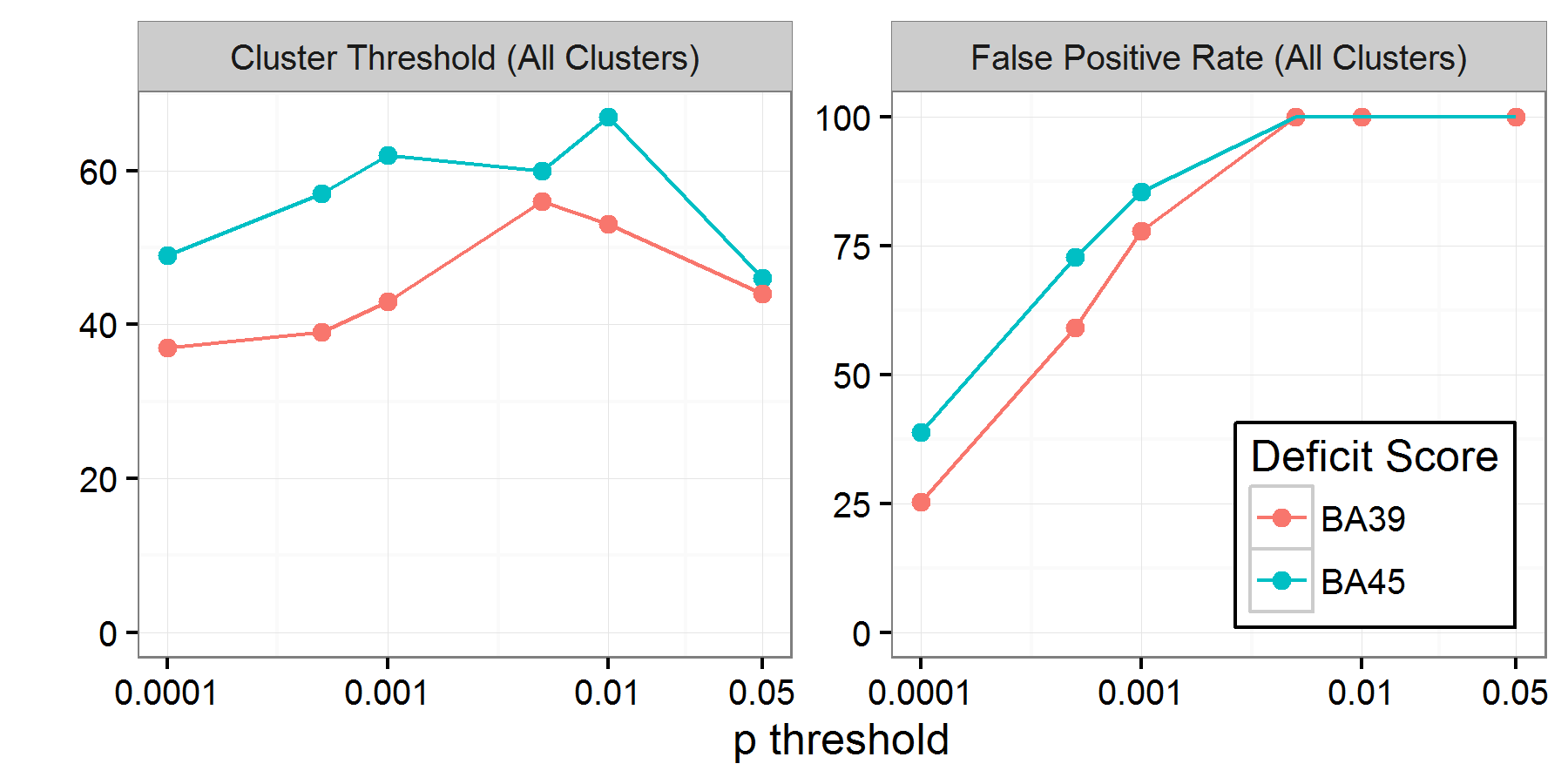}
\caption{Left panel: Cluster size thresholds based on all clusters at each p-threshold. Right panel: Percent of permutations with clusters larger than the cluster size threshold in the left panel (i.e., the false positive rate) at each p-threshold.}
\label{fig:allclust}
\end{figure}

Based on our interpretation of prior work \citep{Pillay2014, Mirman2015a}, in our first analysis, all clusters generated by each permutation were entered into the null distribution of cluster sizes for computation of the 95\% threshold. The relationship between p-threshold and cluster size threshold is shown in the left panel of Figure \ref{fig:allclust}. Surprisingly, the positive relationship between p-threshold and cluster size threshold was not found – the critical cluster sizes were approximately the same across all p-thresholds. To examine this further, we computed the proportion of permutations that contained at least one cluster larger than the 95\% threshold in the left panel of Figure \ref{fig:allclust}. This is measure of false positive rate in the sense that the permutations, by definition, have no systematic relationship between deficit score and lesion location. Therefore, any clusters that survive this correction would be false positives. This false positive rate is shown in the right panel of Figure \ref{fig:allclust} and was alarmingly high relative to the nominal rate of 5\%: it was 25\% for the most conservative p-threshold and rose to 100\% for thresholds $\geq 0.005$.

This very high rate of false positives is particularly problematic because the cluster size threshold eliminates small clusters, so the false positives that survive this correction will be large and likely to induce unsuspecting researchers to make strong claims about lesion-deficit relationships. This analysis was motivated by reading the method section of \citet{Pillay2014}, which had led to a secondary analysis reported by \citet{Mirman2015a}, though it is possible that we did not implement their analysis strategy correctly. In their discussion of permutation-based methods for neuroimaging data analysis, \citet{Nichols2002} specified that only the maximum cluster size should be used from each permutation when building the null cluster size distribution, which is analogous to the maximum t-value used for voxel-wise permutation-based correction. In Analysis 2 we re-analyzed the data from the 1000 permutations reported here using only the maximum cluster size.

\subsection{Analysis 2: Maximum clusters}

\begin{figure}[hb]
\centering
\includegraphics[]{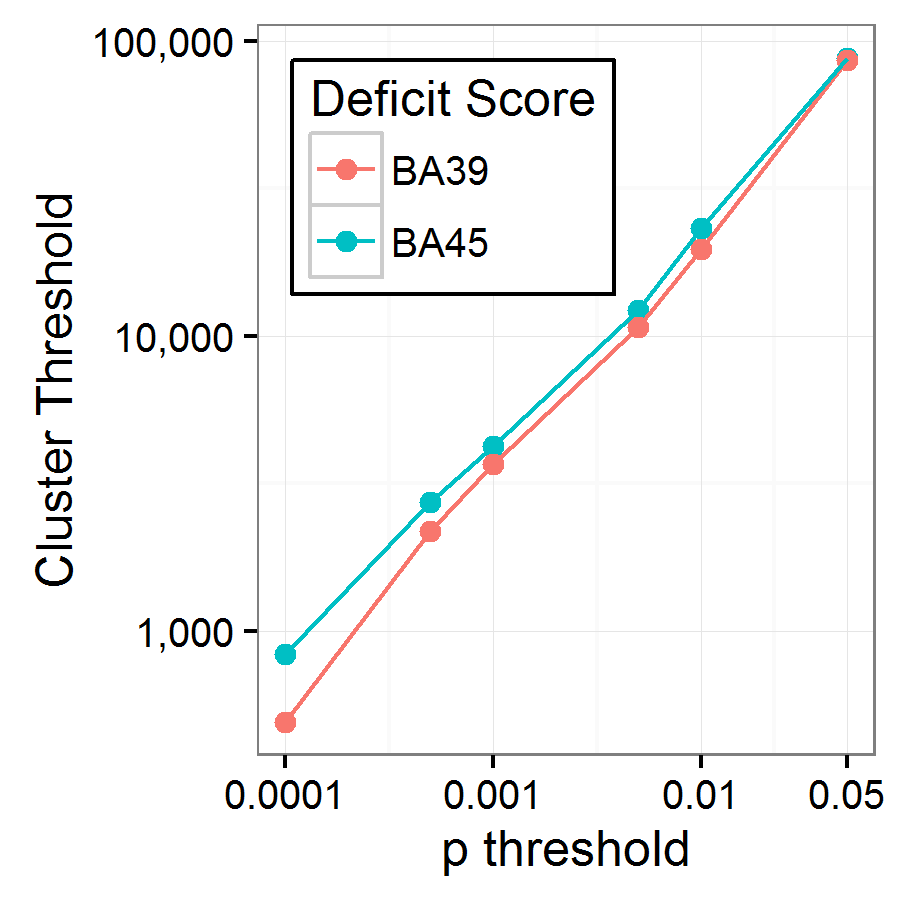}
\caption{Cluster size thresholds based on largest cluster from each permutation at each p-threshold. Note that both axes are logarithmically scaled.}
\label{fig:maxclust}
\end{figure}

The relationship between cluster size threshold (95th percentile of maximum cluster sizes across 1000 permutations) and p-threshold is shown in Figure \ref{fig:maxclust}. This analysis produced the expected positive relationship between p-threshold and cluster size threshold: more permissive p-thresholds allow more voxels into the cluster analysis, thus producing larger clusters. Indeed, the relationship is almost perfectly linear in the log-log plot in Figure \ref{fig:maxclust}. Analysis of false positives is trivial here because, by definition, 95\% of permutations had 0 clusters larger than the cluster size threshold.

The next stage was evaluating how well this method recovers the true neural correlates for each deficit score. It was immediately apparent that only the most conservative p-threshold (0.0001) produced a viable cluster size threshold. The next most conservative threshold (0.0005) produced cluster size thresholds of more than 2000 voxels; at p-threshold 0.01, the cluster size threshold was above 10,000 voxels. Any clusters of that size or larger would not be neuroanatomically specific enough to provide useful insights into lesion-symptom relationships. 

\begin{figure}[ht]
\centering
\includegraphics[width=1.0\textwidth]{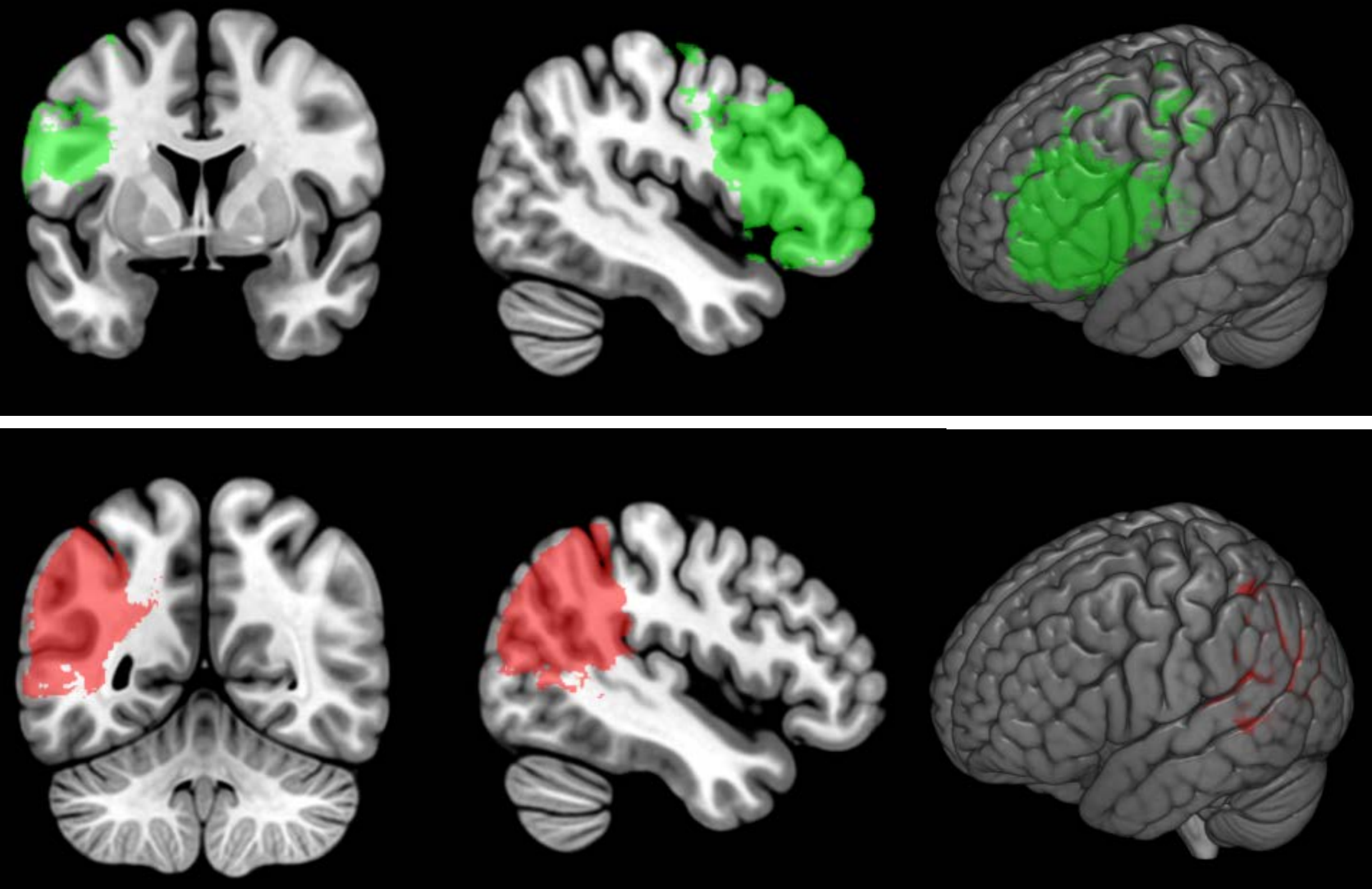}
\caption{Results of VLSM analysis of percent damage to BA 45 (top row) and BA 39 (bottom row). Thresholded at voxel-wise $p < 0.0001$ and permutation-based cluster size corrected.}
\label{fig:BA39BA45}
\end{figure}

Figure \ref{fig:BA39BA45} shows the results of permutation-based cluster size correction (at $p < 0.0001$) for simulated deficit scores of percent damage in BA 45 (top row) and BA 39 (bottom row). It is immediately apparent that the identified region expends beyond the bounds of the true region, covering an area that is perhaps twice the size of the Brodmann Area where percent damage was used as the behavioral score. 

\begin{figure}[ht]
\centering
\includegraphics[width=0.67\textwidth]{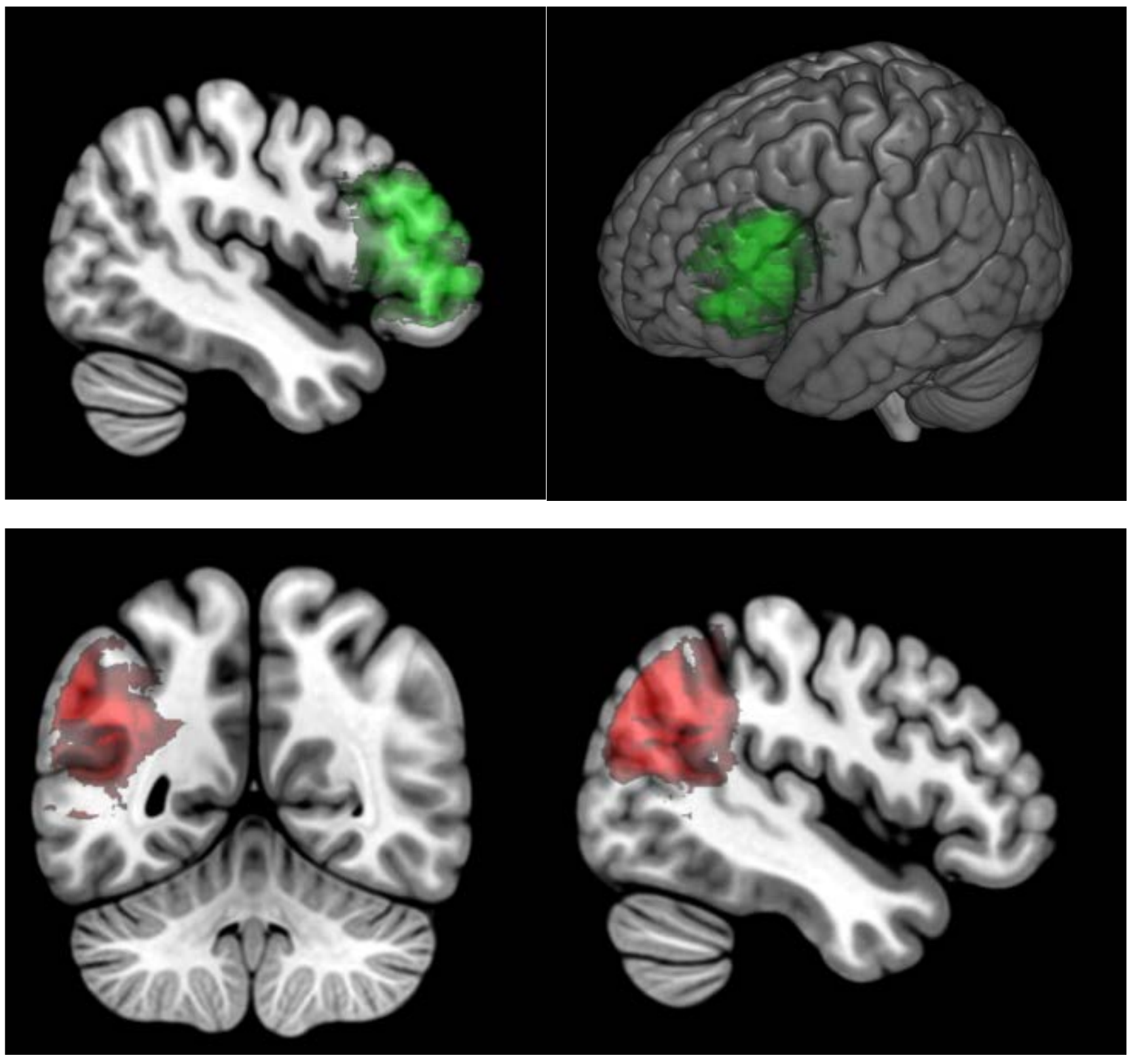}
\caption{Results of VLSM analysis of percent damage to BA 45 (top row) and BA 39 (bottom row). Thresholded using permutation-based FWER at $p < 0.05$.}
\label{fig:BA39BA45FWER}
\end{figure}

For comparison, we used the maximal t-value from the same 1000 permutations to compute permutation-based FWER-corrected $p < 0.05$ thresholds for each of these analyses. The resulting t-thresholds were 5.52 for BA 45 and 5.39 for BA 39 and the thresholded maps are shown in Figure \ref{fig:BA39BA45FWER}. This approach did a reasonably good job of identifying the critical regions.

\section{Discussion and Conclusions}

We explored using a permutation approach for determining a minumum cluster size threshold for statistical correction of VLSM. This approach is adapted from analysis of functional neuroimaging data \citep{Nichols2002} and has been previously used in VLSM \citep{Pillay2014, Mirman2015a}. Using structural lesion data from 124 participants left hemisphere stroke, we constructed deficit scores using percent damage in BA 45 and BA 39. These behavioral scores were used to assess the permutation-based cluster size correction method on three criteria (1) correctly detecting the relationship between damage to BA 45 or BA 39 and the corresponding deficit score, (2) falsely detecting clusters in the permutations (which, by definition, have no lesion-deficit relationships), and (3) incorrectly detecting voxels outside the critical BA 45 and BA 39 regions. 

Our first discovery was that it is absolutely critical that only the largest cluster from each permutation is included in the null distribution of cluster sizes -- using all clusters produces an absurdly high rate of false positive clusters in the permuted. In fact, this was a \emph{re}-discovery because \citet{Nichols2002} specified that only the largest cluster should be used, though it is not clear whether this was the case in prior applications of this method to VLSM \citep{Pillay2014}. 

When only the largest cluster was used, the false positive rate was controlled (i.e., 95\% of all permutations had no clusters that passed the cluster size threshold) and the critical BA regions were correctly identified. However, the supra-threshold clusters extended well beyond the boundaries of the correct BA regions. This pattern suggests that permutation-based cluster size correction can correctly reject no-signal cases but is insufficiently spatially specific when a true relationship exists. That is, when there is no consistent lesion-deficit relationship, this method should produce false positive clusters at the rate specified by the permutation threshold: at cluster threshold $p < 0.05$, there is less than 5\% chance of detecting a cluster if no lesion-symptom relationship exits. However, if a lesion-symptom relationship does exist, this method will detect the correct region but spatially contiguous regions will also be included in the critical cluster. This spill-over effect may lead to incorrect interpretation of the results, so a better method is needed. 

An important limitation of the present analyses is that our simulated deficit scores (percent damage in particular Brodmann Areas) implemented very strong and consistent lesion-symptom relationships. These strong relationships may have contributed to the spill-over effect for the correction based on cluster size and to the good performance of permutation-based FWER. A weaker or more variable lesion-symptom relationship is more likely to be missed by the very conservative FWER strategy and it is not clear whether the cluster size strategy would work better in such cases. In sum, there continues to be a need for statistical correction methods for VLSM that facilitate cluster-level interpretation and provide a more effective balance between false positive and false negative results.

\section*{Acknowledgements}
We thank Yongsheng Zhang for sharing analysis code, and Branch Coslett and the Laboratory for Cognition and Neural Stimulation for helpful discussions. We are particularly grateful to Dr. Myrna F. Schwartz and her research team for sharing the anatomical data that made these analyses possible and for her comments on an early draft of this report. This research was funded in part by Drexel University and National Institutes of Health grant R01DC010805 to DM.

\bibliography{ClusterPermutationMS.bbl}

\begin{thebibliography}{}

\bibitem[Bates et~al., 2003]{Bates2003}
Bates, E., Wilson, S.~M., Saygin, A.~P., Dick, F., Sereno, M.~I., Knight,
  R.~T., and Dronkers, N.~F. (2003).
\newblock {Voxel-based lesion-symptom mapping.}
\newblock {\em Nature Neuroscience}, 6(5):448--450.

\bibitem[Bennett et~al., 2009]{Bennett2009}
Bennett, C.~M., Wolford, G.~L., and Miller, M.~B. (2009).
\newblock {The principled control of false positives in neuroimaging.}
\newblock {\em Social, Cognitive, and Affective Neuroscience}, 4(4):417--422.

\bibitem[Genovese et~al., 2002]{Genovese2002}
Genovese, C.~R., Lazar, N.~A., and Nichols, T.~E. (2002).
\newblock {Thresholding of statistical maps in functional neuroimaging using
  the false discovery rate.}
\newblock {\em NeuroImage}, 15(4):870--878.

\bibitem[Kimberg et~al., 2007]{Kimberg2007}
Kimberg, D.~Y., Coslett, H.~B., and Schwartz, M.~F. (2007).
\newblock {Power in voxel-based lesion-symptom mapping.}
\newblock {\em Journal of Cognitive Neuroscience}, 19(7):1067--1080.

\bibitem[Lichtheim, 1885]{Lichtheim1885}
Lichtheim, L. (1885).
\newblock {On aphasia}.
\newblock {\em Brain}, 7:433--484.

\bibitem[Mirman et~al., 2015a]{Mirman2015a}
Mirman, D., Chen, Q., Zhang, Y., Wang, Z., Faseyitan, O.~K., Coslett, H.~B.,
  and Schwartz, M.~F. (2015a).
\newblock {Neural Organization of Spoken Language Revealed by Lesion-Symptom
  Mapping}.
\newblock {\em Nature Communications}, 6(6762):1--9.

\bibitem[Mirman et~al., 2015b]{Mirman2015b}
Mirman, D., Zhang, Y., Wang, Z., Coslett, H.~B., and Schwartz, M.~F. (2015b).
\newblock {The ins and outs of meaning: Behavioral and neuroanatomical
  dissociation of semantically-driven word retrieval and multimodal semantic
  recognition in aphasia.}
\newblock {\em Neuropsychologia}, 76:208--219.

\bibitem[Nichols and Holmes, 2002]{Nichols2002}
Nichols, T.~E. and Holmes, A.~P. (2002).
\newblock {Nonparametric permutation tests for functional neuroimaging: A
  primer with examples}.
\newblock {\em Human Brain Mapping}, 15(1):1--25.

\bibitem[Pillay et~al., 2014]{Pillay2014}
Pillay, S.~B., Stengel, B.~C., Humphries, C., Book, D.~S., and Binder, J.~R.
  (2014).
\newblock {Cerebral localization of impaired phonological retrieval during
  rhyme judgment.}
\newblock {\em Annals of Neurology}, 76(5):738--746.

\bibitem[Rorden and Karnath, 2004]{Rorden2004}
Rorden, C. and Karnath, H.-O. (2004).
\newblock {Using human brain lesions to infer function: a relic from a past era
  in the fMRI age?}
\newblock {\em Nature Reviews Neuroscience}, 5:813--819.

\end{thebibliography}
\end{document}